\documentclass[10pt,a4paper,twoside]{article}
\usepackage{epsfig}
\usepackage{baltlat6}
\usepackage{array}
\usepackage{here}
\begin{document}
\ \
\vspace{0.5mm}
\setcounter{page}{1}

\titlehead{Baltic Astronomy, vol.\,23, 103--109, 2014}

\titleb{STRUCTURAL PARAMETERS OF STAR CLUSTERS: \\ STOCHASTIC EFFECTS}

\begin{authorl}
\authorb{D. Narbutis}{1},
\authorb{R. Stonkut{\. e}}{1},
\authorb{P. de Meulenaer}{1,2},
\authorb{T. Mineikis}{1,2},
\authorb{V. Vansevi{\v c}ius}{1,2}
\end{authorl}

\begin{addressl}
\addressb{1}{Center for Physical Sciences and Technology,\\ Savanori{\c u} 231, Vilnius LT-02300, Lithuania; donatas.narbutis@ftmc.lt}
\addressb{2}{Vilnius University Observatory, {\v C}iurlionio 29, Vilnius LT-03100, Lithuania}
\end{addressl}

\submitb{Received: 2014 June 23; accepted: 2014 July 27}

\begin{summary}
Stochasticity of bright stars introduces uncertainty and bias into derived structural parameters of star clusters. We have simulated a grid of cluster $V$-band images, observed with Subaru Suprime-Cam with age, mass, and size representing a cluster population in the M31 galaxy and derived their structural parameters by fitting King model to the surface brightness distribution. We have found that clusters less massive than $10^4$\,${\rm M}_\odot$ show significant uncertainty in their core and tidal radii for all ages, while clusters younger than 10\,Myr have their sizes systematically underestimated for all masses. This emphasizes the importance of stochastic simulations to asses the true uncertainty of structural parameters in studies of semi-resolved and unresolved clusters.
\end{summary}

\begin{keywords}
galaxies: star clusters: general -- galaxies: individual (M31)
\end{keywords}

\resthead{Structural parameters of star clusters}{D. Narbutis et al.}

\sectionb{1}{INTRODUCTION}

Structural parameters of star clusters have been measured in samples from various environments using different techniques and spatial resolutions aiming to search for evolutionary trends. The most common technique used in extragalactic studies of semi-resolved and unresolved clusters is a 2D structural model fit to the observed surface brightness distribution (see, e.g., Larsen 1999), which assumes a constant stellar number to luminosity ratio in each pixel of the cluster image. However, this assumption is not valid for low-mass clusters, where stochastic bright stars alter cluster's profile.

The problem of how stochasticity affects derivation of cluster's evolutionary parameters (age, mass, and extinction) based on integral photometry has received much attention recently (see, e.g., de Meulenaer et al. 2013 and references therein). Stochastic star-by-star image simulations have been used to derive color-magnitude diagrams of semi-resolved clusters (Larsen et al. 2011), to analyze effects of mass segregation (Ascenso et al. 2009), and to study influence of metallicity and mass segregation on the observed sizes of globular clusters (Sippel et al. 2012). However, a homogeneous analysis of the influence of stochastic effects on the structural parameters of star clusters is lacking.

Here we present a grid of artificial clusters with properties similar to the cluster population in the M31 galaxy observed with Suprime-Cam on the Subaru telescope by Vansevi{\v c}ius et al. (2009). The results could also provide a guidance for more distant clusters observed with the {\it HST}. In Section 2 we present artificial cluster image simulation, in Section 3 model fitting to observations is described, and in Section 4 the influence of the stochastic effects is analyzed and discussed.

\sectionb{2}{ARTIFICIAL CLUSTERS}

We used the ``SimClust'' program by Deveikis et al. (2008) to simulate cluster images. Stellar masses were sampled according to the IMF by Kroupa (2001) and their $V$-band luminosities were computed from stellar isochrones of $Z=0.008$ metallicity by Marigo et al. (2008). Stars were distributed spatially according to a 2D King (1962) model profile with the same probability density (i.e., without mass segregation), which is defined by a central density, $\mu_0$, a core radius, $r_c$, and a tidal radius, $r_t$:
\begin{equation}
\mu(r)=\mu_0\left[{\left({1+\frac{{r^2}}{{r_c^2}}}\right)^{-1/2}-\left({1+\frac{{r_t^2}}{{r_c^2}}}\right)^{-1/2}}\right]^2.
\end{equation}

Distances to the clusters are similar to that of the M31 galaxy. The resolution of the Subaru Suprime-Cam observations with ${\rm FWHM}=3$\,pix of the Gaussian PSF and an image scale of 0.2\,arcsec/pix was assumed. Images were rendered using the ``SkyMaker'' program (Bertin 2009) with a constant sky background of 1000 ADU. To make stochasticity a dominant source of uncertainty over the photon noise, a reduced Gaussian background noise of $\sigma=3$\,ADU was introduced into each pixel.

The following parameters were used to build a grid of artificial clusters: the four ages: 10\,Myr, 100\,Myr, 1\,Gyr, and 10\,Gyr; the six masses: $3\cdot10^3$, $10^4$, $3\cdot10^4$, $10^5$, $3\cdot10^5$, and $10^6$\,${\rm M}_\odot$; the six core and tidal radii combinations: $r_c=0.8$, 1.5, and 3.0, and $r_t=15$ and 40\,pix. The grid covers the star cluster population in M31 studied by Vansevi{\v c}ius et al. (2009), although for completeness extends beyond the derived limits of real clusters. At each node of the grid 100 artificial clusters were simulated.

Examples of the simulated images are shown in top blocks of Figs.\,1--4. Age groups are presented in separate figures, panels correspond to different cluster masses, and six images in each panel show clusters with different input structural parameters. The images are shown with the same limits of pixel values and using {\it asinh} scaling function (Lupton et al. 2004), which enhances low-level features while preserving structure in bright regions.

\begin{figure}[!tH]\vbox{
\centerline{\hspace{10mm}\psfig{figure=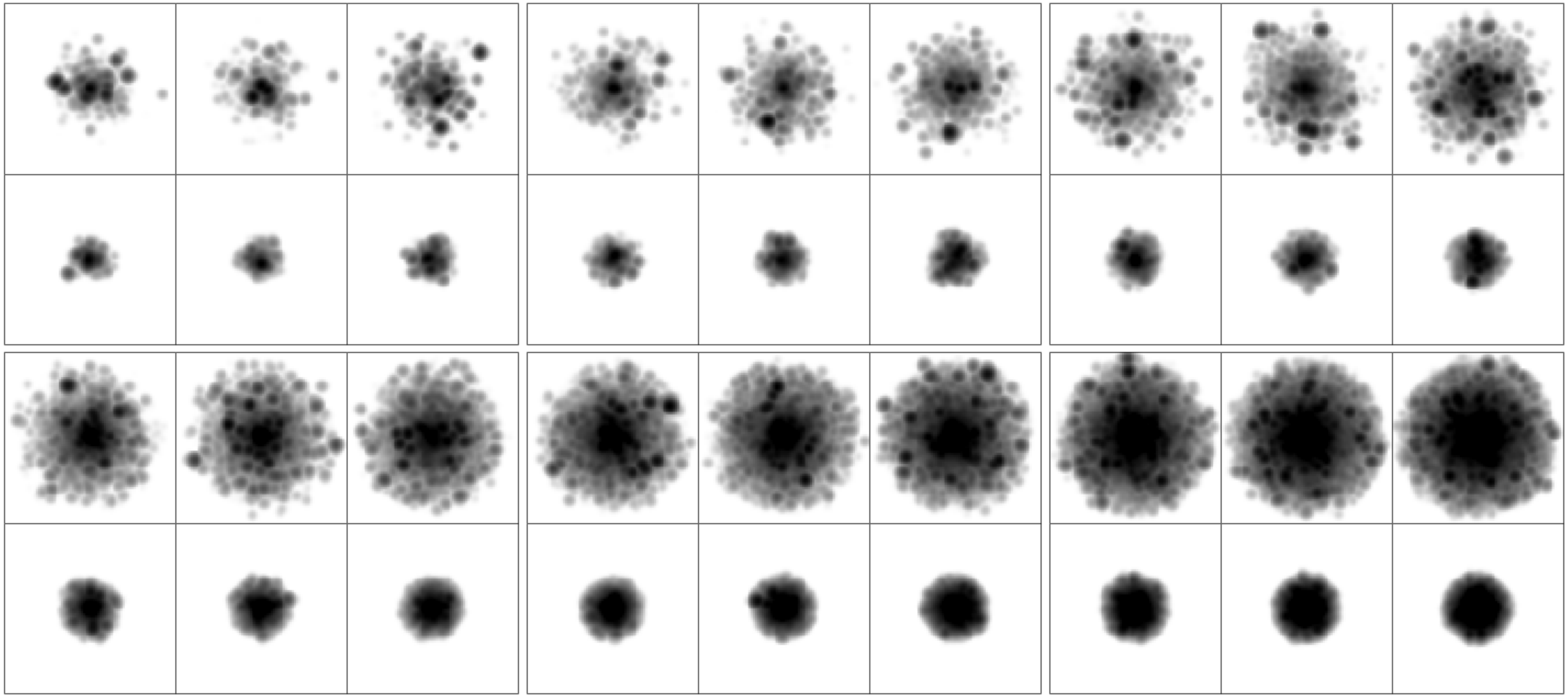,width=110mm}}\vspace{1mm}
\centerline{\psfig{figure=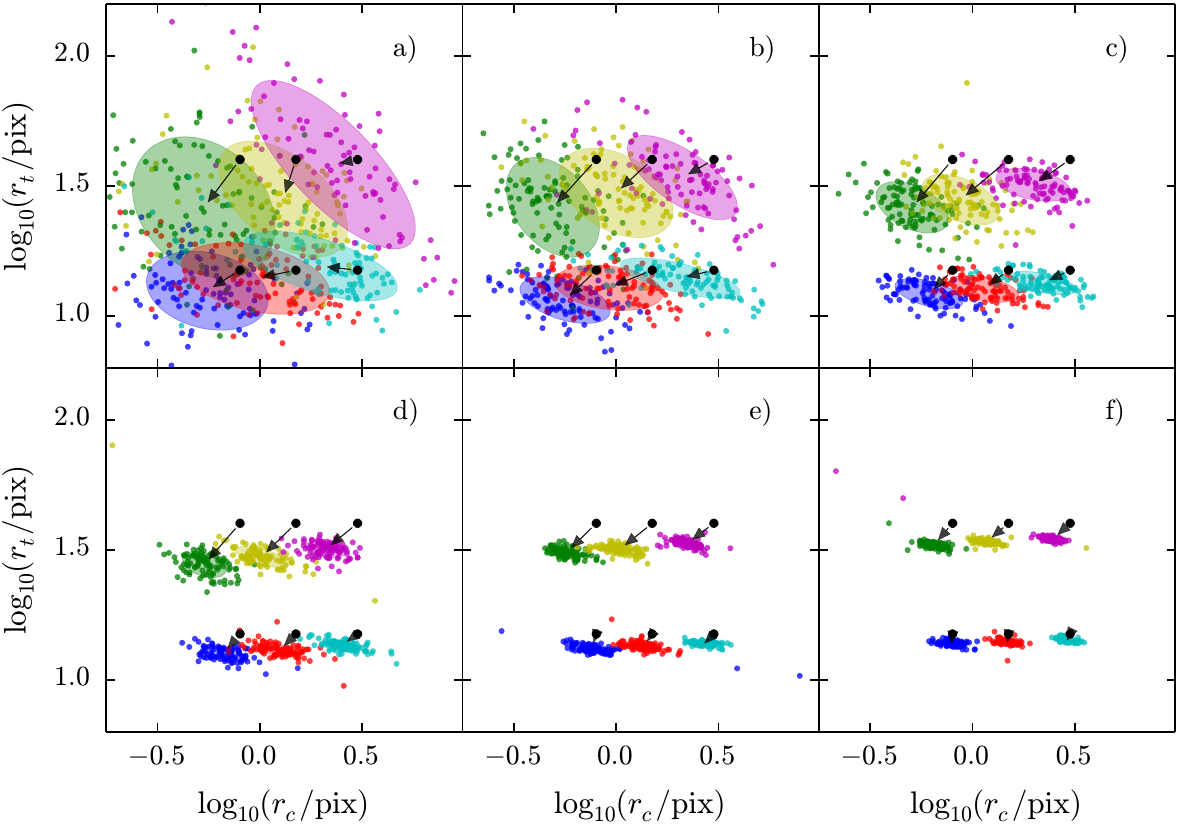,width=120mm}}\vspace{1mm}
\captionb{1}{10\,Myr cluster. Blocks: top -- $V$-band images without sky background, bottom -- King model fit results. Panels in each block show clusters with mass: (a) $3\cdot10^3$, (b) $10^4$, (c) $3\cdot10^4$, (d) $10^5$, (e) $3\cdot10^5$, and (f) $10^6$\,${\rm M}_\odot$. Six example images in each panel of top block correspond to the input structural parameter nodes marked by black dots in the bottom block. Colored dots show the distribution of recovered parameters derived for 100 artificial clusters per each node approximated with $1\sigma$ ellipses, and the vectors indicate bias.}}
\end{figure}

\begin{figure}[!tH]\vbox{
\centerline{\hspace{10mm}\psfig{figure=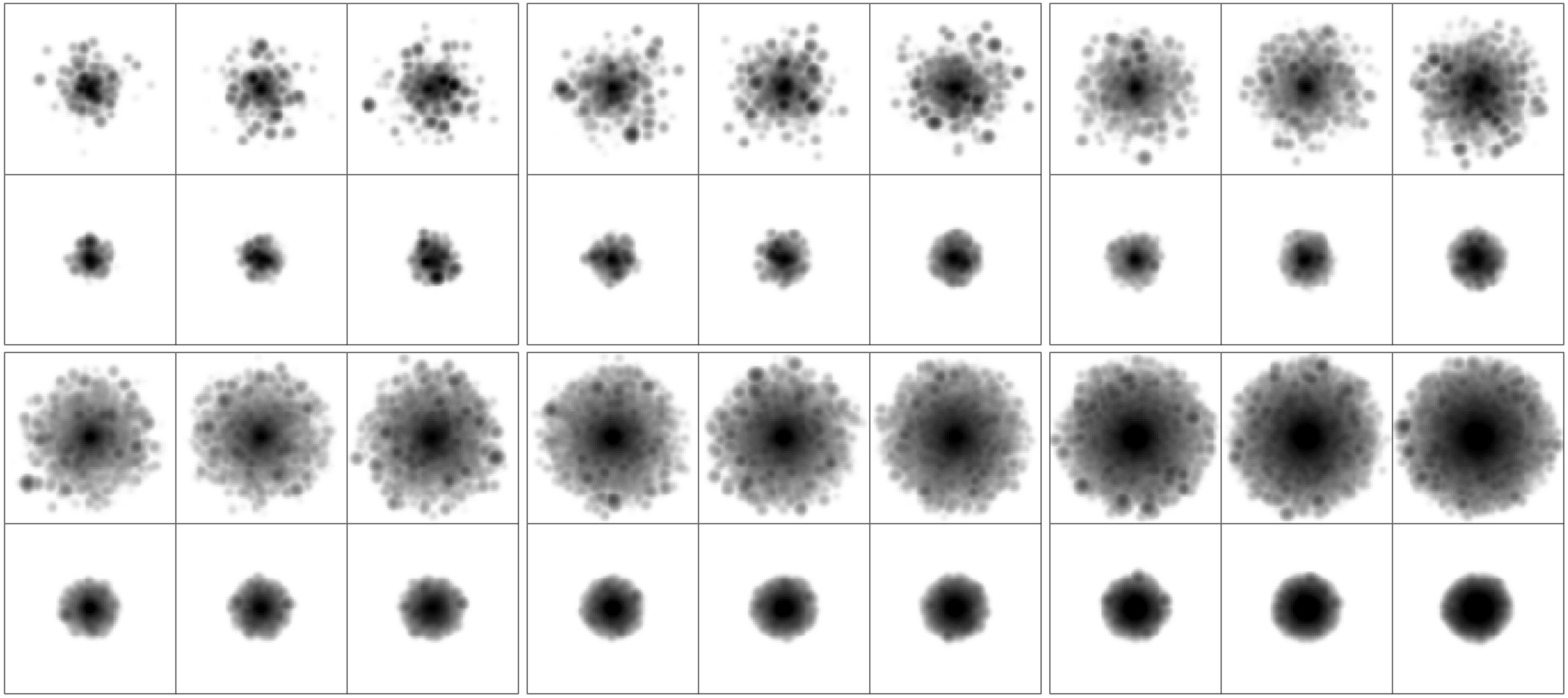,width=110mm}}\vspace{1mm}
\centerline{\psfig{figure=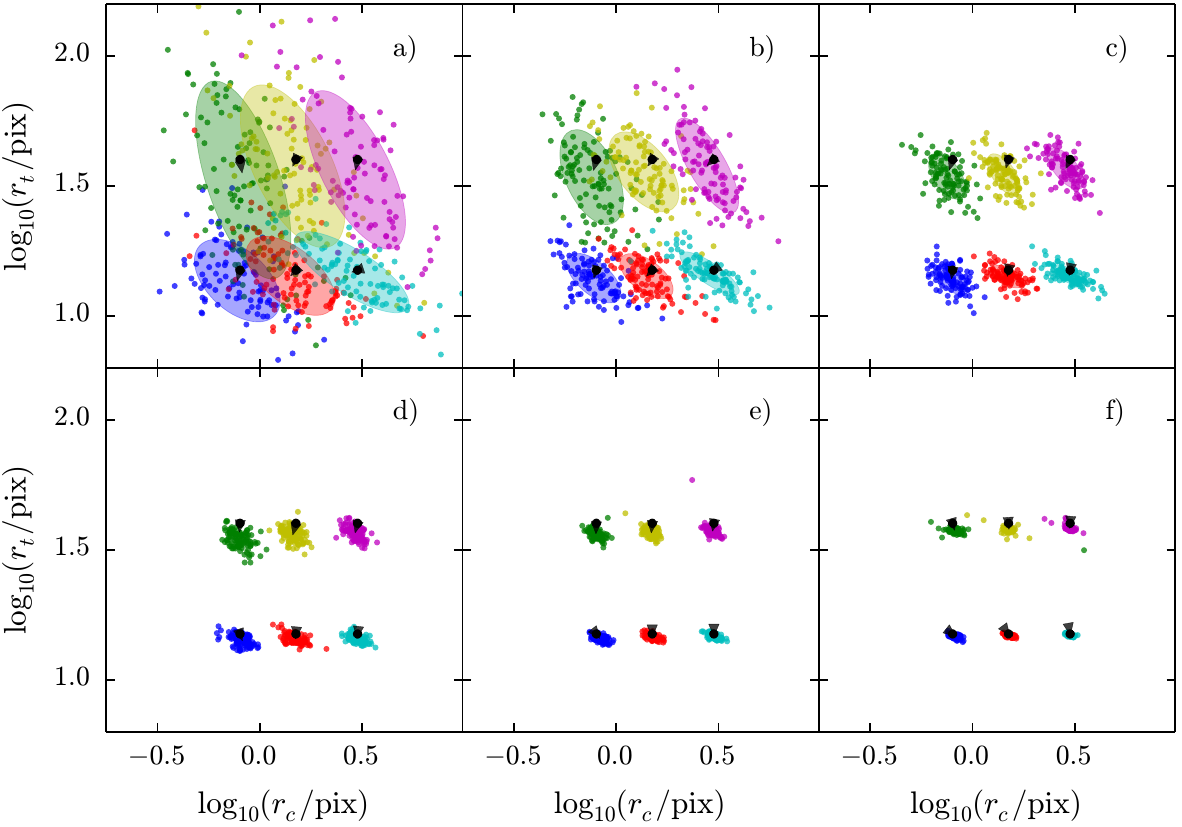,width=120mm,angle=0,clip=}}\vspace{1mm}
\captionb{2}{The same as in Fig.\,1, but for 100\,Myr clusters.}}
\end{figure}

\begin{figure}[!tH]\vbox{
\centerline{\hspace{10mm}\psfig{figure=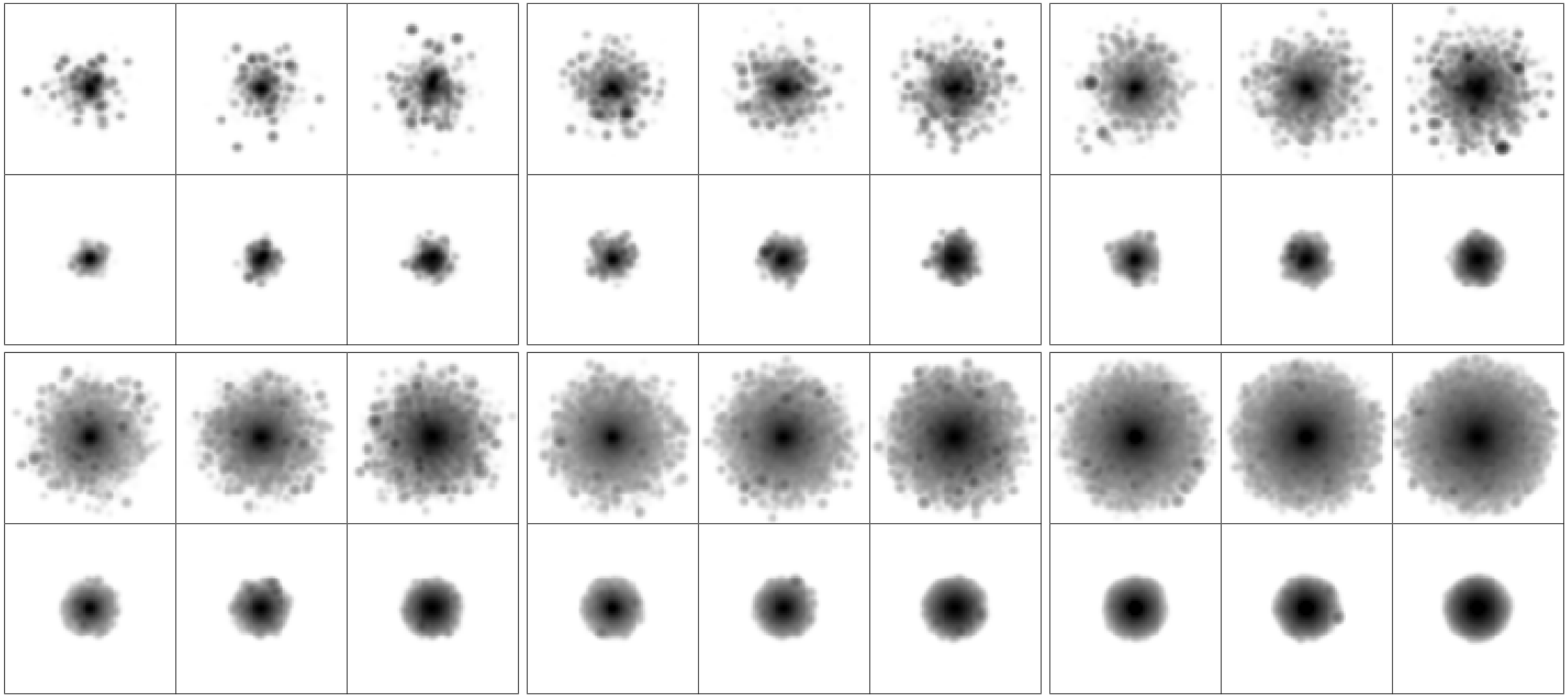,width=110mm}}\vspace{1mm}
\centerline{\psfig{figure=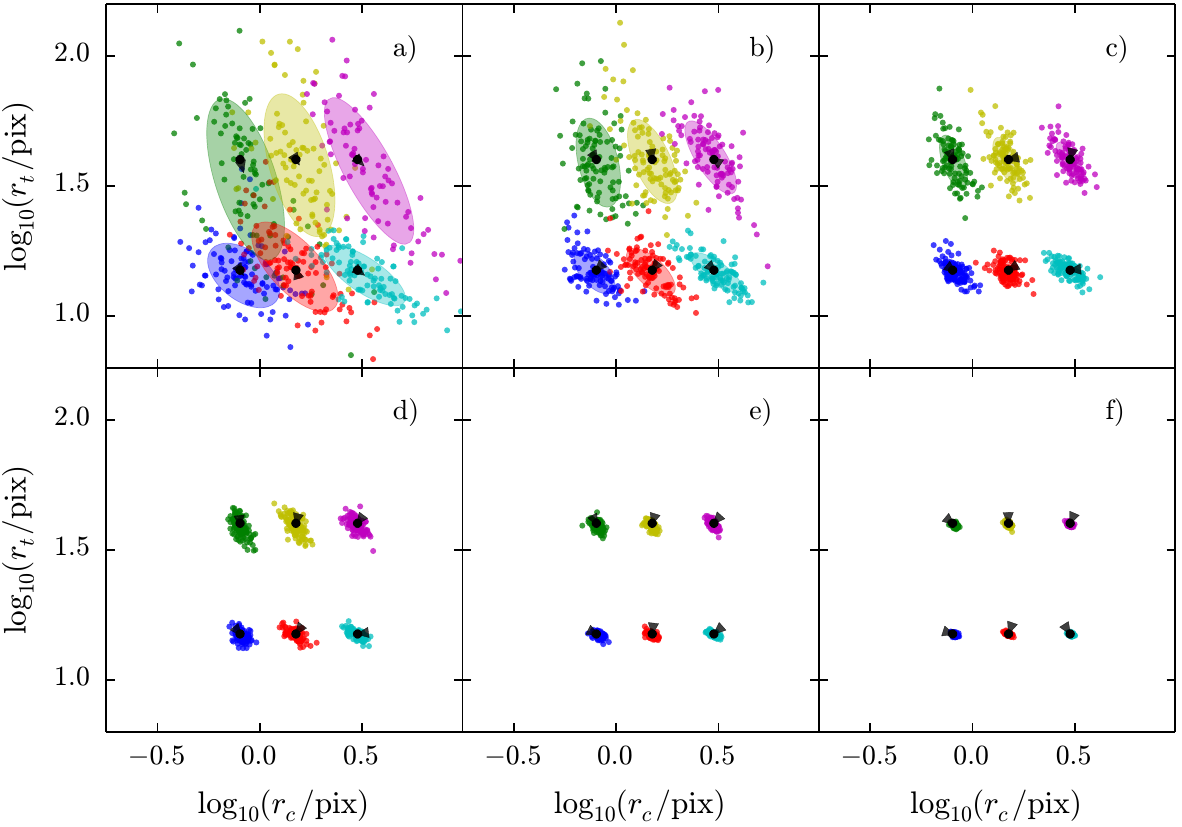,width=120mm,angle=0,clip=}}\vspace{1mm}
\captionb{3}{The same as in Fig.\,1, but for 1\,Gyr clusters.}}
\end{figure}

\begin{figure}[!tH]\vbox{
\centerline{\hspace{10mm}\psfig{figure=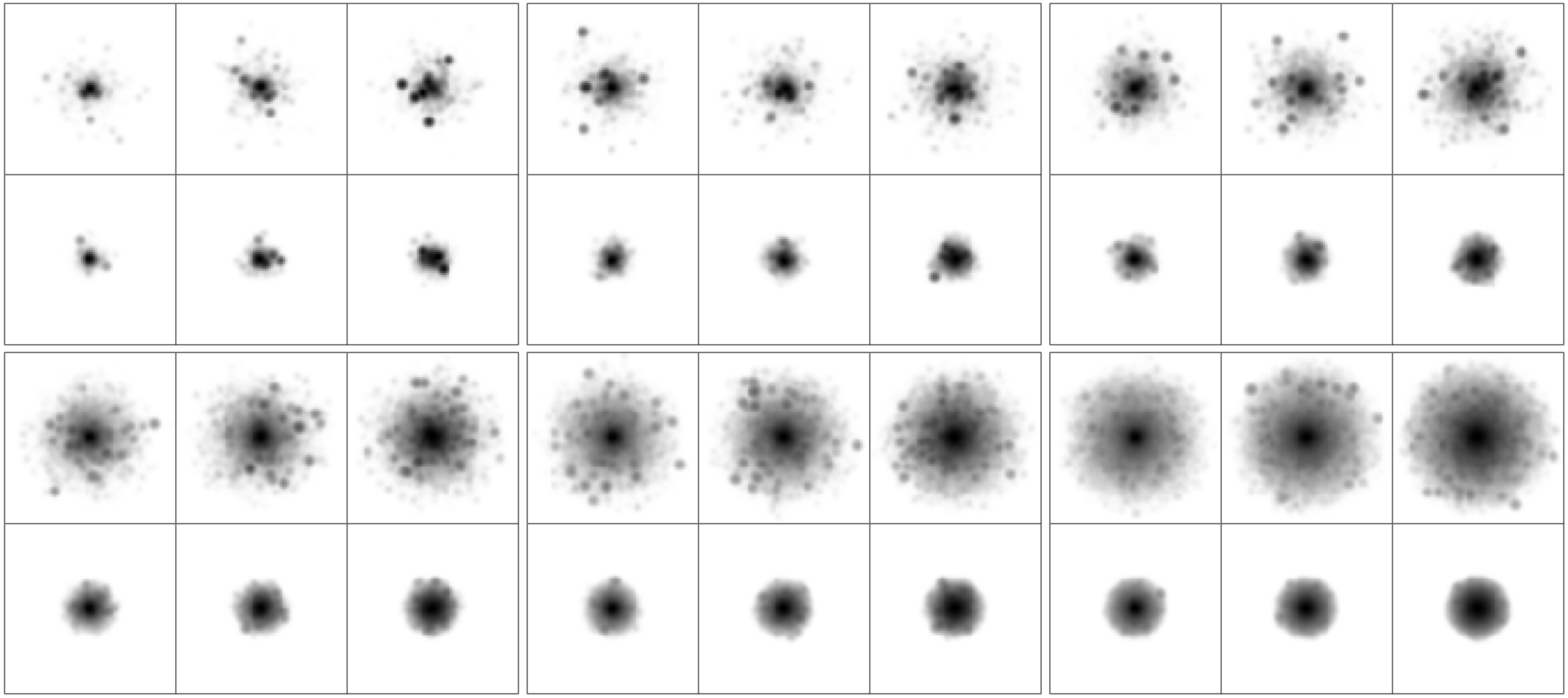,width=110mm}}\vspace{1mm}
\centerline{\psfig{figure=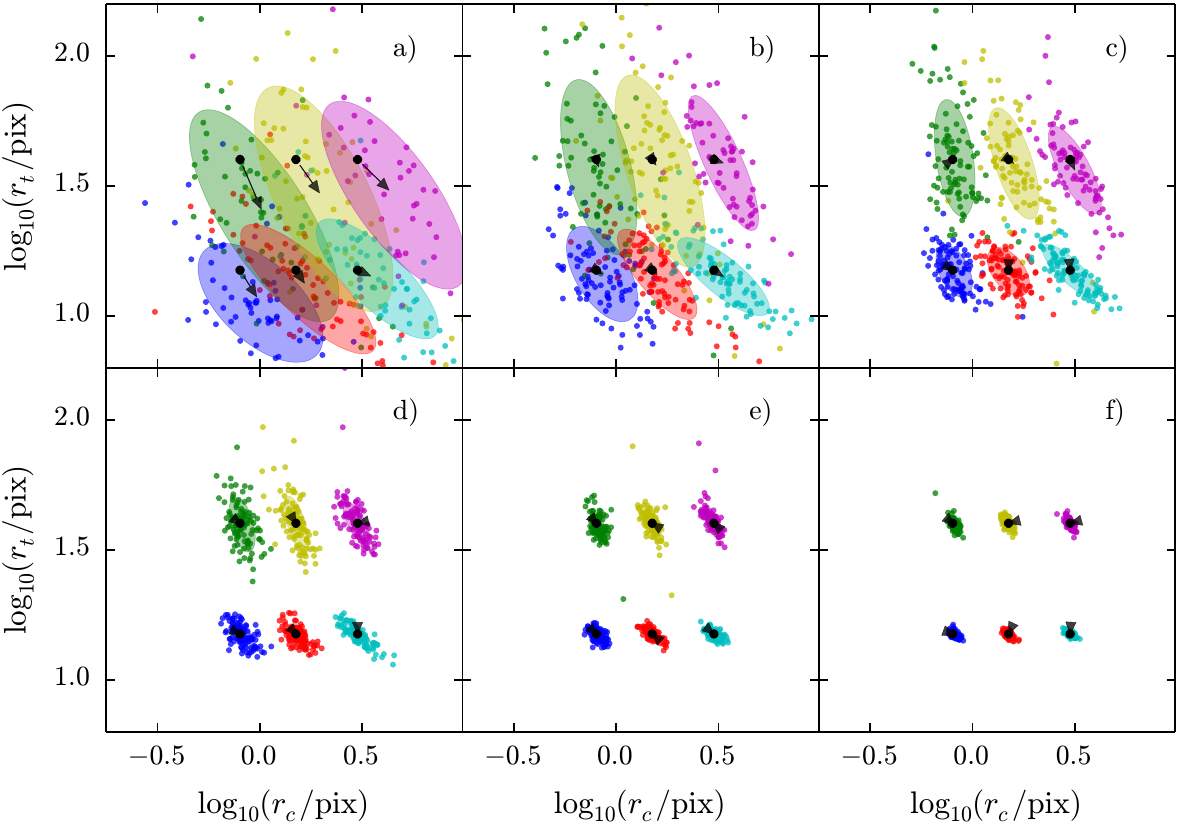,width=120mm,angle=0,clip=}}\vspace{1mm}
\captionb{4}{The same as in Fig.\,1, but for 10\,Gyr clusters.}}
\end{figure}

\sectionb{3}{MODEL FITTING}

We used the ``{\it emcee}'' implementation (Foreman-Mackey et al. 2013) of the Markov Chain Monte Carlo (MCMC) sampler to derive structural parameters of star clusters. We first subtract a constant sky background from an image and then fit a smooth 2D King surface brightness distribution convolved with the PSF to account for the observational effects. The center position of a model is fixed to the input position of a cluster and only the three parameters are fitted: core radius, tidal radius, and total flux.

To initialize the ``{\it emcee}'', we start from the input parameter values of core and tidal radii of the grid node, while flux is set to the integral flux computed for individual cluster. Then the ``{\it emcee}'' samples the parameter space and at each step evaluates the goodness of the model fit, which is a likelihood of the observed image to be generated by a given smooth model and assuming Gaussian noise of the data.

After 3000 steps of a burn-in phase, the maximum likelihood in the parameter space is reached, the subsequent sample of 1000 steps is analyzed and the medians of each parameter are assumed to be the best-fit values. They are displayed as scattered dots in bottom blocks of Figs.\,1--4. We note that the parameter uncertainty reported by the MCMC model fit of each cluster is much smaller than the scatter of the same age-mass-size cluster group due to stochastic effects.

\sectionb{4}{RESULTS AND DISCUSSION}

The bottom blocks of Figs.\,1--4 display results of model fitting for four star cluster ages (10\,Myr, 100\,Myr, 1\,Gyr, and 10\,Gyr, respectively), while the panels correspond to six cluster masses. Six input structural parameter nodes are indicated by black dots in each panel. Color-coded dots show distribution of recovered parameters for 100 clusters per node. Arrows connect input parameters with centers of $1\sigma$ ellipses approximating recovered parameters.

The youngest clusters of 10\,Myr (Fig.\,1) show a significant uncertainty due to stochasticity in both core and tidal radii. As the mass of clusters increases, the relative stochastic influence of the brightest stars becomes smaller (i.e., the surface brightness distribution becomes smoother), therefore, the scatter of recovered core and tidal radii decreases. It is interesting that structural parameters are biased for all masses of young clusters -- their core and tidal radii are systematically smaller, but the systematic shift decreases with increasing cluster mass. However, even for the most massive clusters of $10^6$\,${\rm M}_\odot$ the recovered sizes are smaller than the input values. Therefore, the bright stars significantly alter surface brightness distribution of young clusters and caution should be taken when studying cluster samples to derive evolutionary trends, because of younger clusters appear systematically smaller.

As a relative number of bright stars decreases with cluster's age, 100\,Myr clusters (Fig.\,2) show smaller scatter of parameters, and no systematic shift is observed. Comparing Fig.\,1\,(b) to Fig.\,2\,(b) we see that the uncertainty of core radius becomes smaller while the uncertainty of tidal radius remains the same. Older (1\,Gyr) cluster images (Fig.\,3) are more smooth giving smaller uncertainty. However, at 10\,Gyr (Fig.\,4) clusters show larger parameter uncertainty than at 1\,Gyr, especially for the lowest cluster mass of $3\cdot10^3$\,${\rm M}_\odot$ (Fig.\,4\,a), which have systematically decreased tidal radii. Examination of their images reveals that the reason for this is a lower signal to noise ratio in the outskirts of 10\,Gyr clusters.

In the M31 cluster sample (Vansevi{\v c}ius et al. 2009) objects with the mass $10^4$\,${\rm M}_\odot$ are considered as massive ones, however, for all ages the uncertainty of their structural parameters is non-negligible, and it is much more significant for lower mass ($3\cdot10^3$\,${\rm M}_\odot$) objects, which are more numerous.

\sectionb{5}{CONCLUSIONS}

We have simulated a grid of stochastic cluster $V$-band images, with properties similar to the cluster population in the M31 galaxy observed with the Subaru Suprime-Cam, and derived their structural parameters by fitting a 2D model to the observed surface brightness distributions.

We have found that stochastic effects of bright stars introduce uncertainty and bias into derived structural parameters of star clusters: (1) clusters less massive than $10^4$\,${\rm M}_\odot$ show significant uncertainty in their core and tidal radii for all ages, while (2) clusters younger than 10\,Myr have their sizes systematically underestimated for all masses.

This emphasizes the importance of stochastic simulations to asses true uncertainty of structural parameters in studies of semi-resolved and unresolved clusters when looking for evolutionary trends.

\thanks{This research was funded by a grant No. MIP-074/2013 from the Research Council of Lithuania.}

\References

\refb Ascenso J., Alves J., Lago M. T. V. T. 2009, A\&A, 495, 147

\refb Bertin E. 2009, MmSAI, 80, 422

\refb Deveikis V., Narbutis D., Stonkut{\. e} R., Brid{\v z}ius A., Vansevi{\v c}ius V. 2008, Baltic Astronomy, 17, 351

\refb King I. 1962, AJ, 67, 471

\refb Kroupa P. 2001, MNRAS, 322, 231

\refb Larsen S. S. 1999, A\&AS, 139, 393

\refb Larsen S. S., de Mink S. E., Eldridge J. J. et al. 2011, A\&A, 532, A147

\refb Lupton R., Blanton M. R., Fekete G. et al. 2004, PASP, 116, 133

\refb Foreman-Mackey D., Hogg D. W., Lang D., Goodman J. 2013, PASP, 125, 306

\refb Marigo P., Girardi L., Bressan A. et al. 2008, A\&A, 482, 883

\refb de Meulenaer P., Narbutis D., Mineikis T., Vansevi{\v c}ius V. 2013, A\&A, 550, A20

\refb Sippel A. C., Hurley J. R., Madrid J. P., Harris W. E. 2012, MNRAS, 427, 167

\refb Vansevi{\v c}ius V., Kodaira K., Narbutis D. et al. 2009, ApJ, 703, 1872

\end{document}